\documentstyle[preprint,aps]{revtex}
\begin{document}
\title{ {\bf Faddeev Null-Plane Model of the Proton
\footnote{To appear in "Proceedings VI Hadrons 1998", Florian\'opolis, 
Santa Catarina, Brazil}}}
\author{W. R. B. de Ara\'ujo$^a$, 
J. P. B. C. de Melo$^b$ and T. Frederico$^c$}
\address{$^a$ Instituto de F\'\i sica, Universidade de S\~ao Paulo \\
01498-970 S\~ao Paulo, S\~ao Paulo, Brazil \\
$^b$ Division de Physique Th\'eorique, Institut de Physique Nucl\'eaire,
91406 Orsay Cedex \\  
and LPTPE Universit\'e P. $\&$  M. Curie, 4 Place Jussieu, 75252 
Paris Cedex, France \\
$^c$ Dep. de F\'\i sica, Instituto Tecnol\'ogico da Aeron\'autica, \\
Centro T\'ecnico Aeroespacial, 12.228-900 S\~ao Jos\'e dos Campos,\\
S\~ao Paulo, Brazil}
\date{\today}
\maketitle

\vspace{2 true cm}

\section*{Abstracts}

The  proton is formulated as a relativistic
system of three constituent quarks interacting
via a  zero-range two-body force in the null-plane.
The covariance of the null-plane Faddeev-like equation under kinematical 
front-form boosts is discussed. A simplified three-boson model of
the nucleon wave-function  is obtained numerically.
The proton electric form-factor 
reproduces the experimental data for low
momentum transfers and  qualitatively 
describes the asymptotic region.

\section{Introduction}

In this work, we present a calculation of the proton electric form-factor  
($G_E(q^2)$), using a three quark Faddeev wave-function in the null-plane
\cite{1}. 
The null-plane wave-function is obtained from the solution of the Faddeev 
equation  with zero-range force acting between the constituent quarks. 
The totally symmetric 
spatial part of the wave-function is obtained numerically in a three-boson 
calculation. We use the covariance of the  proton model under kinematical
front-form boosts to calculate $G_E(q^2)$, and compare our numerical results
with the avaliable experimental data. $G_E(q^2)$ scales with $q^4$ in the 
asymptotic region and  describes the data for low momentum transfers. 
In the model  both  ingredients  are together:  
relativistic constituent quark and a contact interaction. We test the 
totally symmetric spatial part of the proton wave-function with these two 
minimal ingredients in a calculation of the proton eletric form-factor. 
The confinement has  no explicit role in the model but is buried in the 
effective degrees of freedom. Relativistic dynamics is required by the  mass
of the constituent quarks and the size  of the nucleon wave-function. 
The reason for choosing a three boson dynamics is to just concentrate on
the  
specific form of the spatial wave-function, and obtain insight in to it
without the complications arising with the use of the spin in the 
light-front. Our numerical results indicate that explicit 
confinement
of the constituent quarks is not essential for a good fit of the electric 
form-factor.

\section{The Model}

The  three-boson Faddeev equation  with a pair-contact
interaction in the null-plane, has been discussed 
in detail in Ref. \cite{3}

The null-plane is defined  by $x^+ \ = \ x^o+x^3 \ = 0 $, and
the kinematical momenta for each particle are given by
$k^+ $ and $\vec k_\perp $. 
It is usual to introduce the momentum fraction for each particle
in a given system, $x \ = \ k^+/P^+$, where
$P^+$ is  the total + momentum component.

 The bound-state two-body null-plane 
wave-function for a constant vertex is given in Ref.\cite{4}.
The three-quark bound-state null-plane wave-function, for the
contact interaction, is constructed in terms of the
Faddeev components of the vertex, $ v(x, \vec k_\perp) $, as:
\begin{eqnarray}
\Psi (x_1, \vec k_{1 \perp}; x_2, \vec k_{2\perp})
= \frac{ v (x_1, \vec k_{1\perp}) + v (x_2, \vec k_{2\perp})
+ v (x_3, \vec k_{3\perp})}
{\sqrt{x_1 x_2 x_3} ( M^2_n \ - \ M^2_0) } \ ,
\end{eqnarray}
where $M_n$ is the nucleon mass, and $M_0$  the free three-quark mass . 
Each quark has  momentum
fraction $x_j$ and transverse
momentum $ \vec k_{j\perp}$ (j=1,3),  satisfying
$ x_1 + x_2 + x_3 \ = \ 1 $ and $ \vec k_{1 \perp} + \vec k_{2\perp}
+  \vec k_{3\perp} \ = \ 0$ in the nucleon center of mass.

The null-plane Faddeev equation is covariant under
kinematical front-form boosts, as it should be, and the 
the Faddeev component of the vertex transforms under such boosts
as
\begin{eqnarray}
v'(x',\vec k'_\perp ) = v (x, \vec k'_\perp - 
 \vec P_{n\perp} \  x ) \ .
\end{eqnarray}
The prime indicates quantities  in the new frame, 
where the nucleon has transverse momentum  $ \vec P_{n\perp} $.

\section{ Proton electric form-factor}

The proton electric form-factor is identified with the 
three-boson bound-state electromagnetic form-factor.
The use of bosons instead of fermions, is just due
to the fact we are modeling the totally symmetric spatial
part of the nucleon wave-function.
As the wave-function
is totally symmetric under particle permutation,  we can reduce
the electric form-factor to the term in which only particle 3 absorbs the
virtual photon:
\begin{eqnarray}
G_E(q^2) = \int dx_1 dx_2 d^2 k_{1\perp} d^2 k_{2\perp} 
 \Psi^f (x_1,\vec k_{1 \perp}; x_2, \vec k_{2\perp})
 \Psi^i (x_1,\vec k_{1 \perp}; x_2, \vec k_{2\perp}) \ ,
\end{eqnarray}
where $x_3 = 1 - x_1 - x_2$. The transverse momentum
of particle 3 is different for the initial and final wave-functions.

The initial and final proton wave-functions 
in the Breit-frame are given
in terms of the center-of-mass wave-function according
to the boost transformation of the vertex, Eq.(2),
We observe that $G_E(q^2)$ is invariant under
frame transformations related to the Breit-frame by  kinematical
front-form boosts.

{\bf Results.} Our dynamical null-plane model 
of the proton wave-function requires 
two paramaters as input, the constituent quark mass ($M$)
and the meson mass ($\mu$). We restricted $ M $ and $ \mu $ so that
the nucleon mass is 938 MeV. 
The values of $ q^4 G_E(q^2) $ for squared momentum transfers 
below 6 GeV$^2$
are compared with the experimental data of Ref. \cite{5}
in Figure 1. We observe
a qualitative agreement with the avaliable data, 
for  $ \mu / M $ = 1.8 and 1.95 . 
The non-confining nature of our wave-function  
is the reason for the flatness of the product $ q^4 G_E(q^2) $ at high
$q^2 $. Models of the nucleon with confinement 
yield a form-factor which
decreases quickly in the asymptotic region \cite{3},
or they obtain a good agreement with the asymptotic data at
the expense of a small constituent quark mass . 

\vspace{0.5 cm}
 
This work was supported by Conselho Nacional de 
Desenvolvimento e Pesquisa - CNPq, 
Coordena\c{c}\~ao de Aperfei\c{c}oamento de Pessoal 
de N\'\i vel Superior - CAPES and Funda\c{c}\~ao de Amparo a 
pesquisa do Estado de S\~ao Paulo - FAPESP. J. P. B. C. de Melo is a 
FAPESP-Brazil fellow (proc-97/1302-8).

\begin{figure}[h]
\begin{center}
\vspace{15cm}
\caption{Proton $ q^4 G_E(q^2) $ 
for  $q^2 \ < \ 6 \ GeV^2$.The solid line
is the results for $\mu/M=1.95$ and the dashed line for $\mu/M=1.8$}
\includegraphics{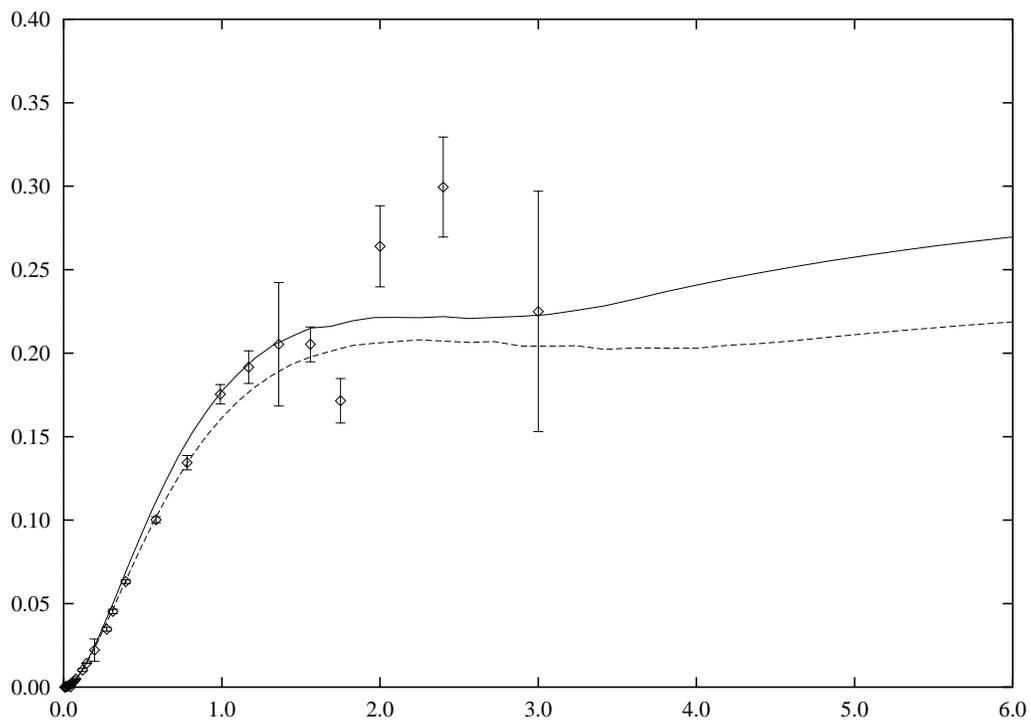}
\end{center}
\end{figure}


\begin{references}

\bibitem{1} W.R.B.de Araujo, J.P.B.C de Melo and T.Frederico, 
Phys.Rev.{\bf C52}, (1995) 2733.

\bibitem{3}Z.Dziembowski,Phys.Rev.D{\bf37},778(1988)

\bibitem{3} T. Frederico,Phys.Lett.B {\bf 282},409(1992)

\bibitem{4} M. Sawicki, Phys.Rev.D {\bf 46}, 474(1992)

\bibitem {5} G. Hohler et. al. Nucl.Phys B {\bf 144},505 (1976)

\end{references}
\end{document}